# *Self-repairing Classification Algorithms for Chemical Sensor Array*

Gabriele Magna[1], Corrado Di Natale[2] and Eugenio Martinelli[1]

1. Department of Electronic Engineering, University of Rome Tor Vergata, via del Politecnico 1, 00133, Roma, Italy
2. Laboratory of Artificial Sensor System, ITMO University, Kronverskiy Prospekt, 49, 197101 Saint Petersburg Russia

Corresponding Author:

Gabriele Magna
Email: magna@ing.uniroma2.it
Post-Doc
Department of Electronic Engineering
University of Rome Tor Vergata
via del Politecnico 1
00133, Roma, Italy

**Abstract**

Chemical sensors are usually affected by drift, have low fabrication reproducibility and can experience failure or breaking events over the long term. Albeit improvements in fabrication processes are often slow and inadequate for completely surmounting these issues, data analysis can be used as of now to improve the available device performances. The present paper illustrates an algorithm, called Self-Repairing (**SR**), developed for repairing classification models after the occurrences of failures in sensor arrays. The procedure considers replacing broken sensors with replicas and eventually Self-Repairing algorithm trains these blank elements. Unlike the habitual alternatives reported in literature, SR performs this operation without the need of a whole new recalibration, references gas measurements or transfer dataset and, at the same time, without interrupting the on-going procedure of gas identification. Furthermore, Self-Repairing algorithm can utilize most of the standard classifiers as core algorithm; in this paper SR has been applied to k-NN, PLS-DA and LDA as examples. Models have been tested in a synthetic and real scenario considering sensor arrays affected by drift and eventually by failures. Real experiment has been performed with a set of metal oxide sensors over an 18-months period. Finally, the algorithm has been compared with standard version of chosen classifiers (k-NN, LDA and PLS-DA) showing superior performances of Self-Repairing and increasing the tolerance versus consecutive failures.

# 1. Introduction

Classification models establish a correspondence between signals incoming from sensors and chemical stimuli [1]. This relationship is based on data collected in a dedicated stage that occurs prior to the classification process, aka training phase. However, the intensive and/or prolonged use of chemical sensors accentuates aging or poisoning of sensing layer giving rise to the so called short- or long-term drift. Because of these occurrences, the responses of sensors to a same chemical stimulus could change over the time and, as a consequence, become different from the ones recorded during the training phase [2-3]. On account of this discrepancy, drift generally produces a gradual lowering in the performances of standard classification models necessitating the acquisition of new calibration samples to periodically train or correct the classifier. However, collecting a whole new training dataset can be time consuming or costly, like for some application concerning the quality control of air in different environments [4-6] or in medical applications where samples are not easily accessible. In such a case, it is convenient to utilize procedures that minimize the number of calibration samples [7] or improve the representativeness of training data [8-9]. Alternatively, a target volatile compound, usually low costly and easy available, can be periodically measured and used as reference for drift compensation [10-11]. At the same time the lifetime of model can be extended by ad hoc acquisition protocols, like the modulation of temperature in conductometric sensors [12-13] or other suitable parameters [14-16]. Finally, some models implement a closed-loop mechanism where feedbacks induce an updating of classifier parameters to follow the changes in sensor responses [17-18]. The so called adaptive or evolutionary algorithms belong to this category and have shown the capability to compensate the effects due to aging, poisoning [19-23] and malfunctioning [24-25].

In case of extensive use, the recurrent damaging or poisoning of sensing layer could make unusable one or more sensors of the array. Fault is defined as an episodic event where at least one characteristic of the system presents an unusual behaviour whereas in sensor failure this situation is permanent and the element cannot operate anymore [26]. Here we consider the case of anomalies whereby sensor signals stop being correlated with chemical inputs, e.g. when noise overwhelms signals [27]. Robustness versus these events can be achieved monitoring and detecting fault and failure occurrences in sensor responses [28] or using redundant array [29]. In any case, robustness is a temporary condition since it depends on the type of damage and the number of sensors involved. Indeed, any sensor array can counteract only a limited number of faults depending on its initial redundancy, then the need of replacement is unavoidable in long lasting or intensive applications. Whether replacement involve the whole array or whether it concerns only a single sensor, a full recalibration of model is usually required since the signals of replacing elements might be significantly different from the ones before the failure. On this regard references samples can be used to transfer calibration among different set of sensor arrays [30-31] or the intrinsic variability can be included in the training model [32] to avoid the need for a transfer subset of samples.

Recently we investigated an algorithm that provides standard classifiers with both adaptation and robustness to fault. The approach, called UOS (Unsupervised Online Selection of features) was shown to be effective in case of k-NN, PLS-DA, LDA and SVM [33] using an array having a limited number of sensors and without the inclusion of replicas. Usually gas sensor arrays utilized as electronic noses have an intrinsic redundancy due to the cross-reactivity of broad selective sensors [34] then, in this context, replicas are not mandatory to counteract sensor failures. In case of UOS, the feature selection algorithm dynamically selects an optimal and distinct subset of features for each single measurement.

Starting from these outcomes, we develop an algorithm for overcome the limitations of the most of current approaches that deal with sensor array recalibration. The basic idea consists in the use of fault tolerance algorithms since they can provide reliable prediction even when a portion of the array is no longer working correctly. In this context UOS algorithm demonstrates high robustness to failures as shown in the case studies exposed in ref [33].

This work illustrates an algorithm, hereinafter referred to as Self-Repairing Classification (SR) Algorithm, that operates in case of sensor replacement after failure events. SR utilizes the predictions of the part of the array not compromised by fault to calibrate the new blank sensors included as replacement. Instead of discarding the current model after a failure to build a whole new one, Self-Repairing Classification utilizes the information given by the leftover sensors to continue the classification process and, at the same time, to calibrate new sensors for repairing the model. This is possible thanks to the cross-selectivity of the sensor array that, here, represents a necessary condition to create a fault tolerance and auto calibrating architecture. Unlike conventional methods reported in literature, SR does not require a pool of transfer samples, the starting array should contain any sensor replicas and new kind of sensors can be also calibrated in this way. Finally, since the classification model recovers its initial accuracy and redundancy once SR procedure is completed, sensors can be continuously replaced extending the lifetime of array hypothetically into infinity.

## 2. Unsupervised Online Selection of features UOS

Before describing Self-Repairing algorithm, it behoves us to give a brief description of UOS considering that a full description and investigation of algorithm can be found in [33]. UOS presents an initial choice process during a training-like phase and a second routine during the classification phase or test. The training procedure consists in a preliminary reduction of features in agreement with the Fisher Discriminant Score (FDS) calculated along all the features. The main goal of this initial selection is the removal of unnecessary descriptors.

Considering a classification task having N-classes where the j-th class contains $L_j$ samples, then FDS is the ratio between intra-class and inter-class variance and it is estimated as following:

FDS = SB/SW  [1]

where

$$SB = \sum_{j=1}^{N}(\mu_j - \bar{\mu})^2 \qquad [2]$$

$$SW = \sum_{j=1}^{N}\left(\frac{1}{L_j} * \sum_{k=1}^{L_j}(x_{jk} - \mu_j)^2\right) \qquad [3]$$

In Eq.(3), $x_{jk}$ is the k-th sample belonging to j-th class, $\mu_j$ is the mean value of samples of j-th class and $\bar{\mu}$ is the mean value of all the data. For each i-th feature, FDS ratios are estimated considering all possible pairs of classes. The features that have at least one FDS ratio higher than 1, value used as fixed threshold, are selected and used in test.

Data are placed in a training reservoir that will be utilized and updated during the test. Fig.1 illustrates the route of UOS during the test and, in particular, after the presentation of a new sample that has to be classified. Initially UOS selects a subset of features in agreement with the actual sample and data stored in a training reservoir (see Fig.1). Eventually the new classification model built on the downsized pool classifies the sample. At the end the sample is included in the reservoir replacing the oldest template of the same class. The dynamic selection of features grants robustness versus fault or failure occurrences whereas updating the pool of templates improves the performances in case of drift.

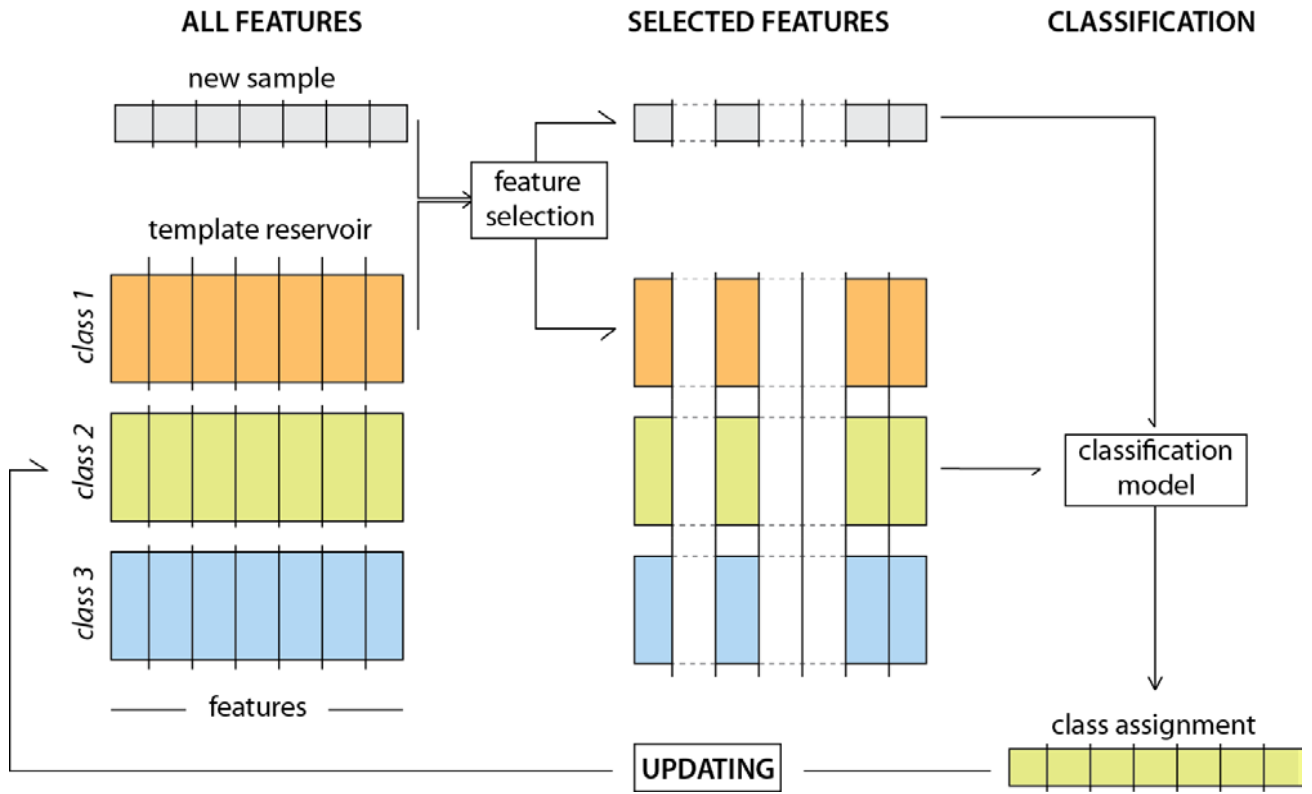

**Figure 1** Graphical depiction of Unsupervised Online Selection of features algorithm routine performed during the test. When a new sample is presented, algorithm performs the feature selection phase. Surviving features are used to build a new classification model using the samples contained in the reservoir. The classification model predicts new sample class that in turn is used to update the template pool.

*Feature selection*

As previously addressed, in UOS the feature selection depends on the sample under classification. Actually a same feature can be useful for discriminating a class from the others but cannot distinguish among the other ones. Furthermore, drift, faults or sensor failures make the information content of features not fixed over the time. Recursive selection aims at identifying which features provide an unambiguous information about the membership of the current sample to a class. In particular, the algorithm selects the descriptors where the test sample can be statistically assigned to only one of the problem classes.

The three criteria used are reported in eq. 4, 5 and 6:

$$prob_{class1} > thresh_1 \qquad [4]$$

$$prob_{class1}/prob_{class2} > thresh_2 \qquad [5]$$

where

$$prob_{class1} = \max_i(prob(x \in class_i)), with\ i \in class\{1,..,N\}$$

$$prob_{class2} = \max_j \left(prob(x \in class_j)\right), with\ j \in class\{1,..,N-1\}: j \neq class1$$

$$Mahal_{class1} * {Mahal_{class1}}/{Mahal_{class2}} < thresh_3 \qquad [6]$$

Where

$$Mahal_{class1} = \min_i \left(mahal(x, \hat{x}_{class_i})\right), with\ i \in class\{1,..,N\}$$
$$Mahal_{class2} = \min_j \left(mahal\left(x, \hat{x}_{class_j}\right)\right), with\ j \in class\{1,..,N-1\}: j \neq class1$$

The first two criteria (eq. 4 and 5) consider the probability of a new sample to belong to the normal distribution built with the templates of a class. This probability is calculated for each class obtaining *prob*$_{class1}$ as the highest probability and *prob*$_{class2}$ as the second highest probability. The two equations state that a feature will be selected only if the sample has high probability to belong to one class distribution and relative low probabilities to the distributions of the other classes. The third equation is the same used in ref. [33] and takes in consideration two factors: the sample must be close to samples of a class (Mahal$_{class1}$) and relative far from the others (Mahal$_{class1}$/Mahal$_{class2}$). It is important to remark that algorithm selects only the features wherein the Class1 indexes resulting from these three criteria are the same. Finally, the same parameters have been used for all the UOS models and datasets as follows: thresh$_1$ is 0.005, thresh$_2$ is fixed to 5 and thresh$_3$ is equal to 0.1. The main influence in the choice of these parameters regards the intrinsic robustness of the in use standard model versus noise or error. For example, LDA and PLS-DA have an intrinsic data dimension compression that should allow to set less constrictive conditions whereas k-NN model requires stricter thresholds. In spite of this, we prefer to choice the same set of parameters for all the cases.

### 3. Self-Repairing classification algorithm (SR)

One of the major benefit of Self-Repairing classification algorithm (SR) is that it performs the calibration of new sensors without interrupting the classification procedure. As a consequence, it can be used to replace a sensor after a failure or just to expand/modify the array configuration. Even though the SR procedure can be combined to large assortment of models addressed by adaptive and fault tolerance properties, coupling SR with UOS, above described, allows to implement the algorithm virtually to any standard classifiers. Here we utilized k-NN, PLS-DA and LDA as an exemplifier subset.

SR algorithm starts once a new sensor is included in the array. The new element measurements are acquired, labelled using the predictions of the existing model (see Fig 2A) and collected in a pool. This reservoir is considered full when, for each class, the number of collected samples is higher than a fixed threshold. Eventually the new pool is included in the training reservoir of UOS model.

*Self-Repairing algorithm*

The first step of SR algorithm consists in identifying the presence of failures during the test. Detection of malfunctioning can be performed by a plethora of algorithms, for example it is possible to utilize a statistical approach [35], monitor the changes in descriptors [36] or using multivariate analysis procedures [37-38]. Here anomalies are supervisedly detected to focus results and investigation on the properties of SR algorithm. Two different kinds of anomalies have been considered: temporary and permanent fault. The former case takes account for situation where sensors are not working for a short period of time, after which they return to operate properly. We can anticipate that in this case the results reported in the dedicated section show the recalibration occurs automatically after a transitory period. In the unlucky circumstance of a permanent failure, the malfunctioning sensor has to be removed and replaced by a replica. In this case the Self-Repairing procedure is initialized (see Fig. 2C) gathering a pool of templates for the new sensor. As every training pool, this reservoir needs an X-block, containing the samples/measurements, and Y-block, containing the labels. X-block consists in the features extracted by the responses of the replacing sensor during the test. Y-block is formed by the predictions of UOS algorithm based on the remaining sensors of array since the broken sensor was removed. SR proceeds until the pool of templates belonging to the new sensor contains the same number of measurements for each class of the reservoir as the UOS model had. In other words, after replacement SR lasts until the reservoir of template in UOS is completely renewed in order to have the same set of measurements both for the replacing sensor and the residual array. Under this condition, the pool gathered by SR can be merged with the UOS reservoir. Eventually replacing sensor features are included in the UOS classification model.

A critical point in Self-Repairing approach is the intrinsic dependence of calibration procedure on the accuracy of predictions made by UOS since these latter are used as labels for the samples of the new sensors. For this reason, the fault tolerant attitude of UOS model is crucial for the final performance of Self-Repairing algorithm. Accurate details about fault detection and pool sizes are given in the experimental section.

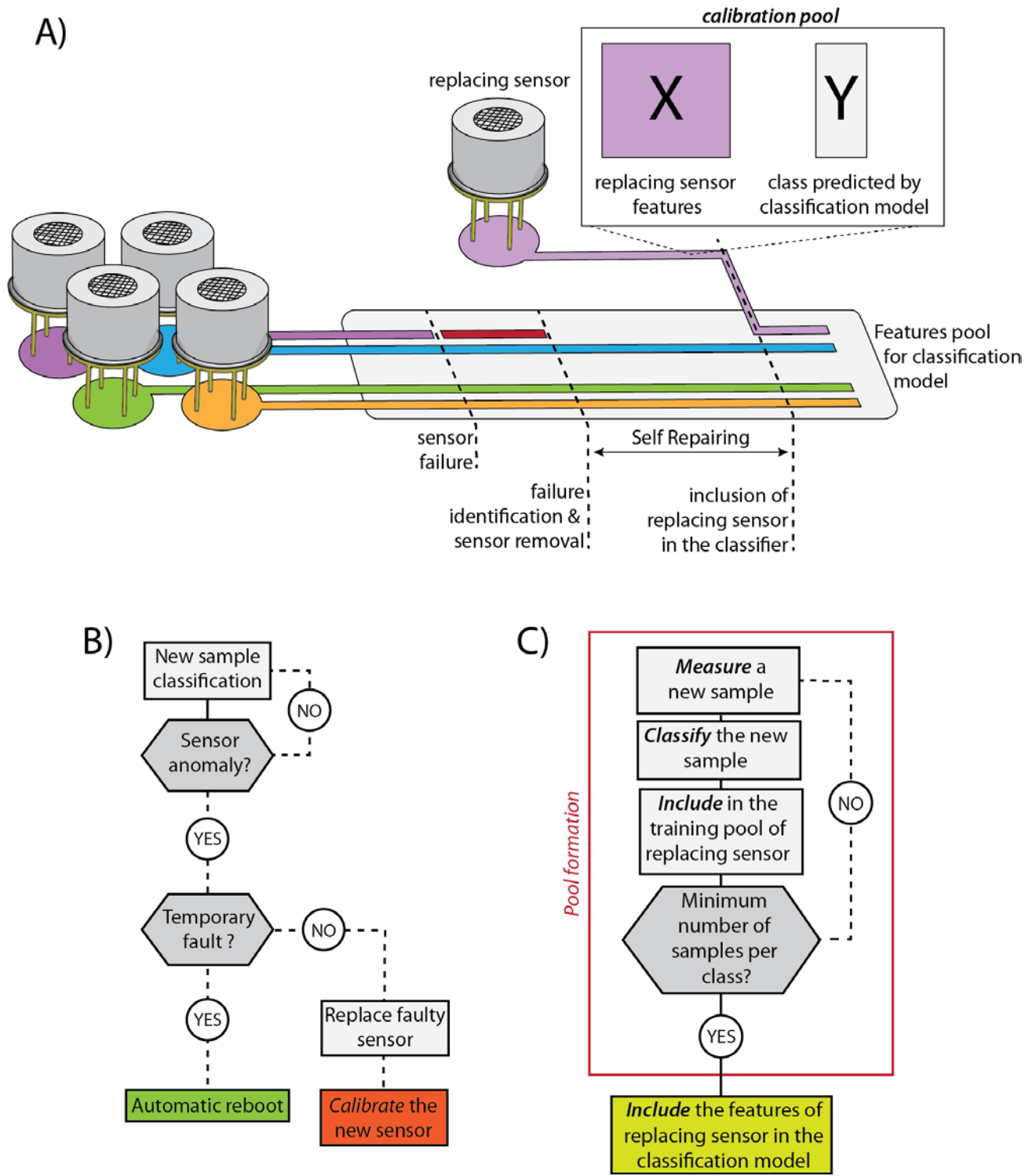

**Figure 2.** A) illustrates the case of sensor replacement carried out by SR algorithm. After the failure identification the malfunctioning sensor is removed and substituted by a new one of the same kind. Measurements of the new sensor are stored until a predefined number. Eventually the features pool used for building the classification model. B) the panel illustrates the decision making process implemented in Self-Repairing algorithm. In case of temporary fault, the system automatically proceeds to recalibration. If a sensor failure is detected a new calibration is performed. C) shows the implementation of Self-Repairing in UOS considering a new sensor replacement.

## 4. Dataset

Self-Repairing algorithm has been validated with a synthetic and an experimental dataset. Both datasets are affected by drift and have sensor replica to allow replacement.

*Synthetic Dataset*

The synthetic dataset has been elaborated following the procedure described in Ref. [39]. A three-class problem is faced with an array of five virtual sensors affected by a common linear drift. Training data consist in 60 measurements equally subdivided in the three categories and test set is made of 1200 samples. To simplify the investigation and performance of SR algorithm, samples are alternating between the three classes during the test. However, a more complex situation concerning the class samples presentation has been considered in the real experiment dataset.

Figure 3A shows the whole dataset projected along the first two Principal Components. As it clearly showed data are affected by severe drift and classes 1 and 2 are partially overlapped over the time. Fig 3B shows the timeline of class presentation over the 1260 samples composing the synthetic dataset. Synthetic replicas are included in the dataset considering that the responses of these replacing sensors can differ up to the 20% from the original to simulate fabrication dissimilarities. Selectivity of replicas is consistent with the copied sensor one. Please note that no information about time is given about the training pool.

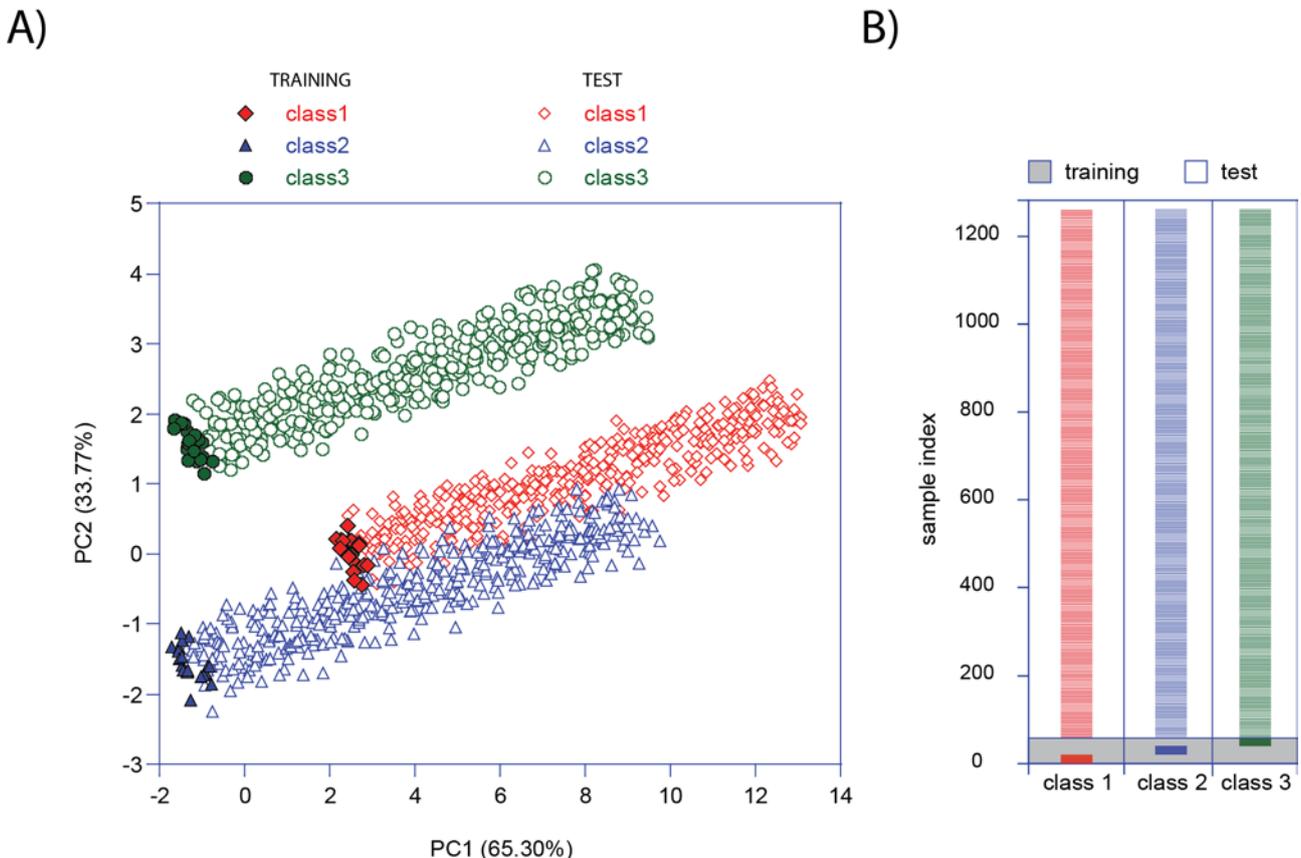

**Figure 3**. A) PCA plot (PC 1-2 projection) of synthetic subset of data used to validate the Self-Repairing algorithm. B) illustrates the time course of experiment considering the chronological order of measurements.

*Experimental dataset*

The experimental dataset is extracted from an extensive set of data available online [40]. The full dataset contains 13910 measurements from 16 chemical sensors applied to the discrimination of 6 gases presented at variable concentration over a period of 36 months. Two of the six compounds started to be measured at a later stage. The 16 sensors are 4 replicas of 4 different commercial Metal Oxide Semiconductor Gas Sensors (MOx) produced by Figaro Inc. (TGS2600, TGS2602, TGS2610 and TGS2620). For each exposure eight feature per sensor is calculated including both steady state and dynamic state descriptors. Two descriptors have been used as steady state features: the change and the relative change of sensor resistance between immediately before and at the end of the gas exposure. On the other hand, transient behaviour of sensors is described by exponential moving average (EMA) both during adsorption and desorption phase. Three different tau have been taken into account resulting in six dynamic features for each measurement.

The subset of data used with the SR algorithm comprises 300 measurements corresponding to three volatile compounds: 50 ppm of acetaldehyde, 250 ppm of ethylene, and 1 ppm of toluene. The first 60 samples were used to train the algorithm while the remaining 240 simulated the normal use of the sensor system to identify unknown samples. This last set is used to test the algorithm.

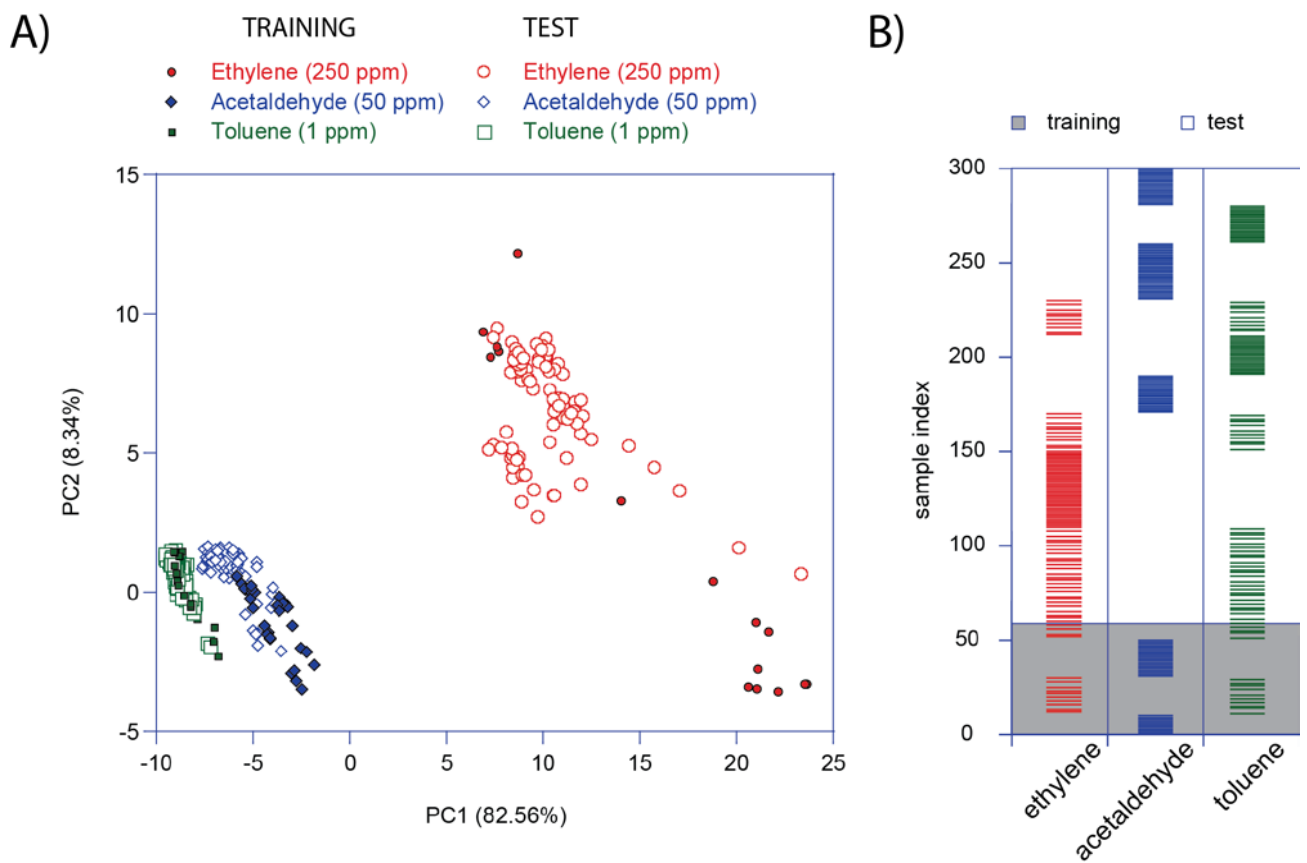

**Figure 4.** A) PCA plot (PC 1-2 projection) of experimental subset of data used to validate the Self-Repairing algorithm; B) illustrates the time course of experiment considering the chronological order of measurements.

Fig 4 A shows the scores plot the PCA of the training data. Albeit sensor drift is less evident in this case, the temporal sequence in Fig. 4B highlights the unbalance class presentation. This condition can strongly affect adaptive classifiers and it is surely challenging for the SR algorithm

since it needs to collect a sufficient number of samples of all the classes before to introduce the new model in the UOS algorithm.

*Sensor failures*

Sensor failures are virtually induced on the sensor array following the scheduling reported in Fig 5A and 5B for the synthetic and experimental dataset respectively. The synthetic dataset situation considers a long lasting experiment where 4 sensors break in sequence. Faults are set after 200, 400, 600 and 800 sample indexes. In order to improve the statistical significance of results, the simulations have been performed 100 times selecting each time a new random sequence of faulting sensors. Since the real experiment has both lesser duration and an unbalanced occurrence of classes, only one fault occurs after 10 samples. In this case fault is induced on all the features linked to the malfunctioning sensors. Even if 8 features are extracted from each sensor, it is worth to remember that the initial feature selection can reduce the number of features actually included in the UOS model. As for synthetic dataset, the sensor that faults are arbitrary selected over 100 simulation runs. Finally, the replica used for the SR algorithm is randomly selected among the three unused sensors of the same kind of faulting sensor.

As previously mentioned, the algorithm robustness towards failures is validated considering two different scenarios where sensor fault is virtually induced on elements of array, similarly to the case reported in [33]. The first case simulates a completely loss of chemical sensitivity, then the sensor responses are set to zero for all the classes considered in the task. In case of random fault, the responses of sensor arbitrarily oscillate in the usual working range. This case could reflect high noise conditions. Fig 5 C and D report these two scenarios applied to the experiment dataset.

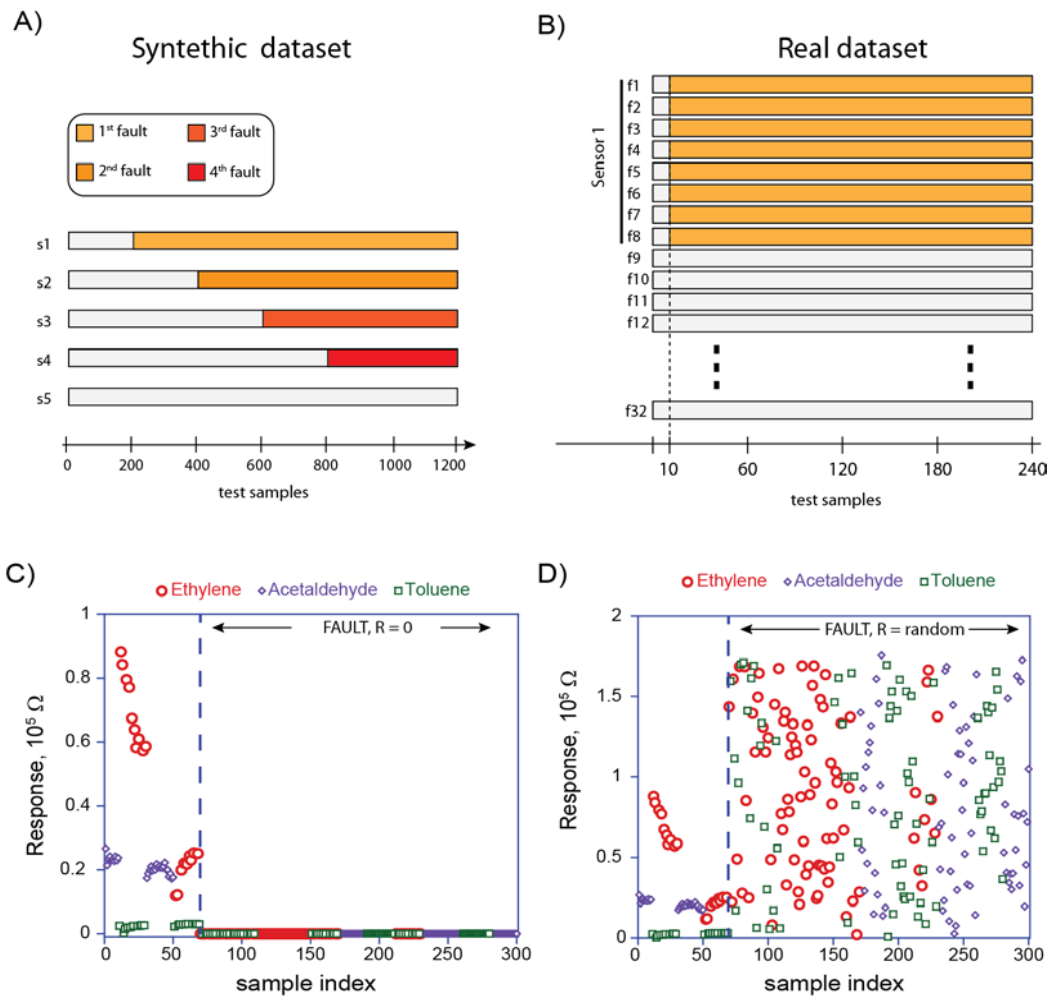

**Figure 5.** A) Failure sequence represented in the synthetic dataset scenario. Four random selected sensors fail consecutively. B) In the experimental dataset case, all the features belong to faulty sensor simultaneously fail after 10 samples in test. C) Example of zero-response failure induced on the resistance responses of a sensor used in the experimental dataset. D) Example of random-response failure induced on the resistance responses of a sensor used in the experimental dataset.

## 5. Results and discussion

The performances of Self-Repairing are evaluated using three standard classifiers: k-NN, LDA and PLS-DA. Results are related to one hundred runs for each algorithm.

### 5.1 Self-Repairing

*Synthetic Dataset*

Standard classifiers achieve very poor classification rate in case of synthetic dataset because of the extensive drift simulated. k-NN (k=3), PLS-DA and LDA obtain 47.2%, 44.3% and 49.5% respectively. Since the inadequate results obtained, standard classifiers will not be considered in case of faulting scenarios. Implementing UOS algorithm greatly improves these initial scores up to almost perfect classification rates, as shown by Fig. 6A. A different picture emerges when four consecutive faults are induced in the dataset (see Fig. 5A). The sequence of sensor failures is chosen randomly and results of UOS algorithm strongly depends on the combination of sensor malfunctioning as evidenced by the high dispersion of classification rates in Fig 6B. This dependence is accentuated by the choice of not including any sensor replica in the virtual array. In both random and zero scenario, UOS algorithm alone is not able to preserve high classification rates because of the high number of failures, four, with regard of sensor elements, five. On the other hand, the implementation of Self Repairing has two main consequences: increases mean classification rates and reduces the variability in the performances. Indeed, SR results are less sensitive to the sequence of sensor faults. These improvements can be explained by the fact that, since faulty sensors are replaced time by time, the simultaneous failures are reduced to only one. SR procedures is clearly shown by Figure 7. The plot reports the feature selection process occurred in a simulation with UOS algorithm applied to LDA where S3, S1, S4 and S5 consecutively fail. On y-coordinates is represent the index of sample during the test (from 1 to 1200), on x-axis the five sensors are placed. For each sensor, a horizontal bar marker is placed at the indexes where sensor is selected by UOS algorithm. The white boxes indicate the SR operation period. At the end of SR procedure, the new sensors responses are introduced in the UOS model in the place of previous one. From that point the new sensor actively participates to the classification procedure as shown by the rate of selection. It is worth to note that rate of selection of the new sensor is comparable to the original one further confirming the success of training procedure. Finally, the number of measurements required for the SR is close to the number of templates in the pool. The short time required for the calibration results from the high classification rates achieved by model and from the very balanced occurrences of class samples (see Fig. 3B), which is one of the best possible cases for this algorithm.

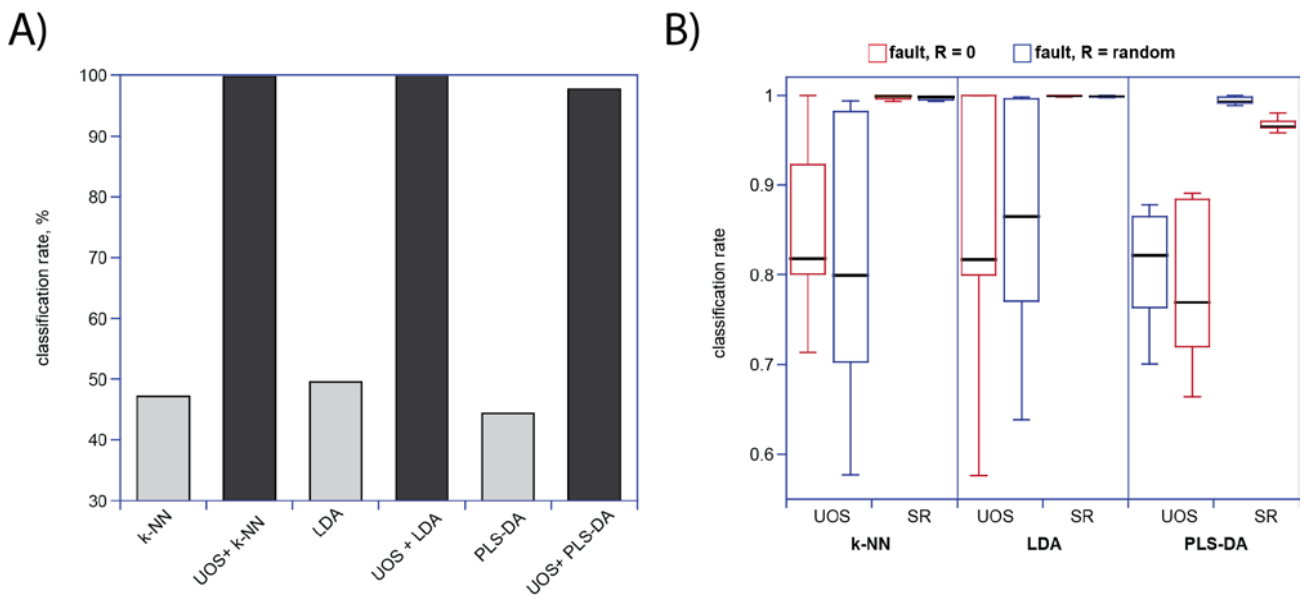

**Figure 6** A) Results obtained by standard classifiers and UOS + standard classifiers in the case of synthetic dataset. B) classification rates achieved by UOS and SR in the case of synthetic dataset affected by two different failure conditions. Statistical variations of model performances are obtained over 100 simulations, each having a random sequence of sensor failures.

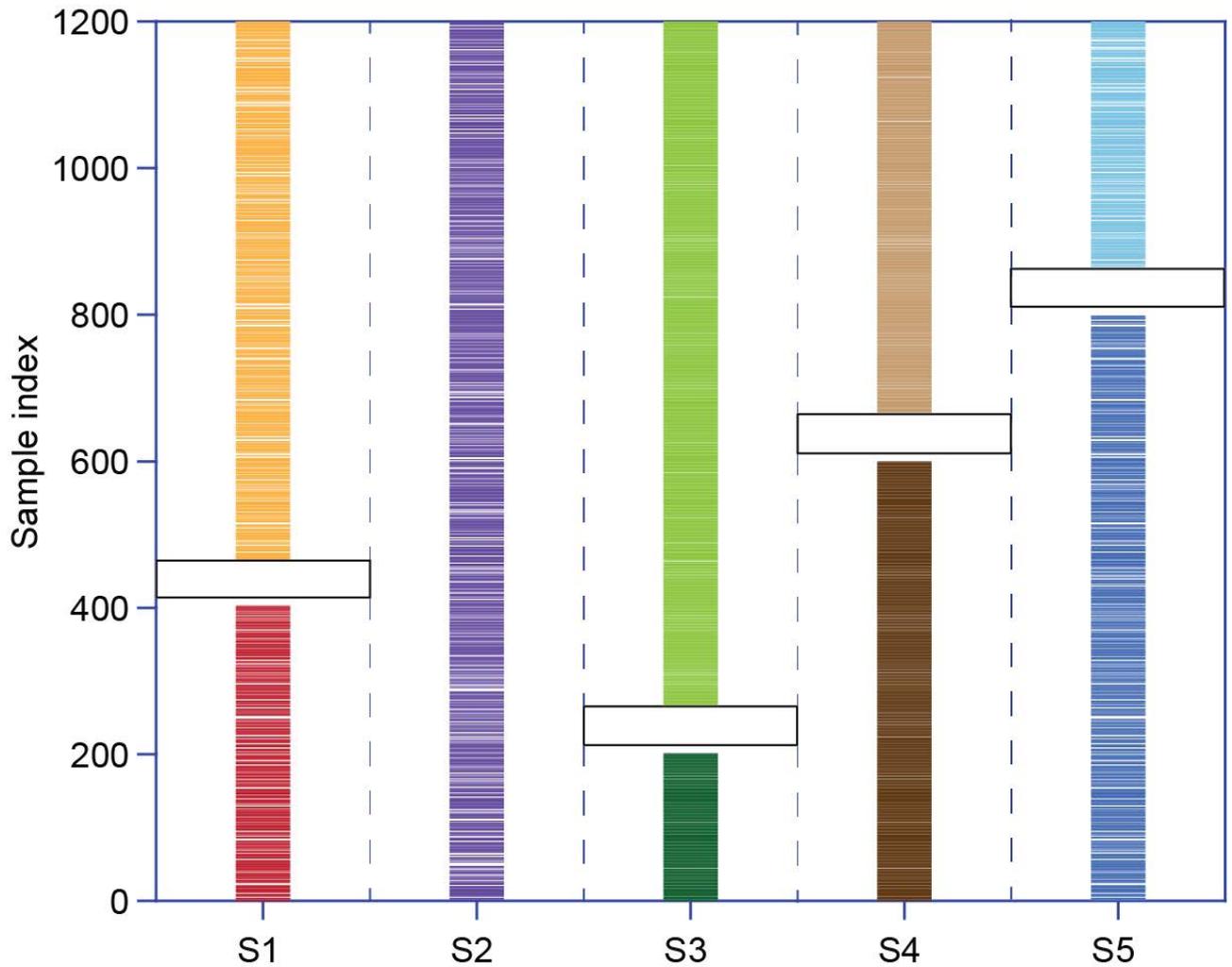

**Figure 7.** Timeline of feature selected in case of one simulation performed using SR + LDA in case of zero-response failures of S3, S1, S4 and S5. White boxes correspond to the periods required for the Self-Repairing procedures. After the fault/failure, the different colour indicates the use of a new sensor.

*Experimental dataset*

Experimental dataset represents a more challenging task for SR and UOS respect to synthetic study case. Albeit drift is less extensive, the occurrence of classes is extremely irregular and not alternated (see Fig. 4B). Indeed, SR algorithm needs a minimum number of samples for each class to gather a new pool of templates then an irregular presentation of data can delay or increase the time required for the calibration. Similarly, the update of latecomer classes is delayed. Finally, in the experimental scenario, replicas are real sensors included from the beginning in the array. Thus replicas are affected by drift, presence of outlier samples and unpredictable noise components. In this situation, the performances of model are more sensitive to the choice of sensor than the synthetic dataset case. For having a comprehensive investigation about SR potentialities, the replacing sensors are selected randomly among the three other sensors of the same kind (as aforementioned four sensors are present in the dataset for each of the four MOS types considered). Fig. 8A reports the results achieved by standard classifiers and UOS algorithm in the case of experimental dataset without failures. Even if the drift is not as extreme as in the synthetic dataset case, UOS implementation produces a statistically significant improvement of performances with respect of standard classifiers. Fig. 8B reports the results relative to sensor malfunctioning occurrences. Since no sensor replicas are included in the classification model, a single failure can strongly influence the accuracy of UOS model. As well as the synthetic dataset case, most of variability in the results shown in Fig. 8B derives from different sensor failures rather than from the iteration of the same fault scenario. As expected, when no replicas are considered, the performances of sole adaptive model are sensitive to which element undergoes to failure. It is worth to note that in spite of this dependence, UOS models achieve high average classification rates (the worst case is 93% for PLSDA in case of random fault). Eventually the inclusion of SR algorithm and the replacement of sensor give rise to a more robust and less aleatory algorithm with respect of both random and zero failure scenarios. Actually Fig 8B show both an increase of mean classification rates and a reduction of interval where results are distributed. Finally, as a further benefit, the system after the new sensor inclusion is robust towards new sensor failure occurrences.

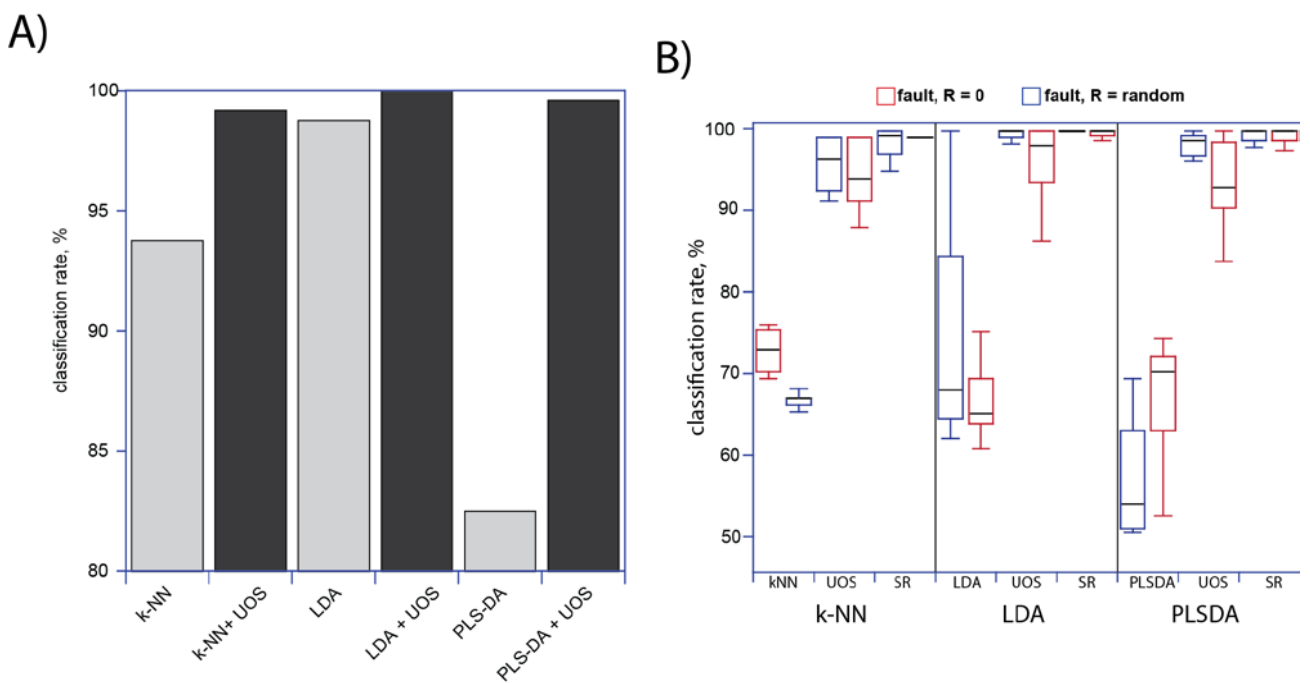

**Figure 8** A) Results obtained by standard classifiers and UOS + standard classifiers in the case of experimental dataset. B) classification rates achieved by UOS and SR in the case of experimental dataset affected by two different failure conditions. Statistical variation of model performances is obtained over 100 simulations, each having a random selection of sensor replica used for replacement.

**5.2 Reboot**

As last study case, it has been investigated the influence of temporary faults on the performance of UOS algorithm both in terms of classification rate and the process of feature selection. Before discussing about temporary fault, let us see what happens in case of permanent failure. Fig. 9A and 9B report the feature selected by UOS algorithm in case of zero-response failure. As a first evidence, the features related to a faulty sensor are not selected by the algorithm. This outcome is on one hand desired since misleading responses are excluded from the classification model and, on the other hand, it provides an indicative clue about the status of sensors. A failure detection algorithm can be developed on the basis of the selection rates of the features corresponding to a same sensor.

Finally, Fig. 9A shows that during the last two faults, the algorithm selects faulted features for a period after the failure occurrence; this behaviour is more pronounced for the sensor 5. The delay in the exclusion of the features coming from a malfunctioning element can be justified by the gradual reduction of the performance of classification model due to multiple faults. Classification errors induce a wrong distribution of classes in the reservoir of templates because of the adaptation mechanism. In turn, the erroneous information about class may promote the selection of not informative features. In this context, Self-Repairing algorithm recovers the information after sensor failures preventing the amplification of errors inside closed loop algorithm like UOS (see Fig.7).

In case of temporary fault, a sensor is not informative for a short temporal window respect to the whole dataset. It has been performed fault scenarios alike the ones depicted in Fig. 5A and 5B, with the difference that faults last only for 15 samples after which the responses come back to original values. Results show that temporary faults do not influence the classification rates with respect to the case when no faults occur (data not shown). This outcome confirms the fault tolerance properties of UOS algorithm, especially in the short-period. A second remark arises from the investigation of feature selection process. Let us consider only the synthetic case since the balanced presentation of classes produce more clear and readable results than the experimental case. Fig. 10 shows the case of S3, S1, S4 and S5 fault sequence at the 200th, 400th, 600th and 800th sample. Faults last for 15 samples and dashed rectangles highlight these temporal windows. Form the plot two distinct situations are evident: in two cases the features of fault sensor are excluded only during the temporal window corresponding to the fault; in the remaining two cases, features are reselected only after a transitory period. Reflecting the points aforementioned, in accordance with the initial class distribution, faulty samples wither can or cannot alter the distributions of classes in the reservoir respect to the actual ones. If distributions are not substantially altered, the feature is immediately reintegrated once fault ends. Conversely the feature starts to be selected again only after the faulty measurements in the reservoir are completely removed thanks to the updating procedure. This last procedure is very similar to an unsupervised SR process where a new reservoir of templates is gathered.

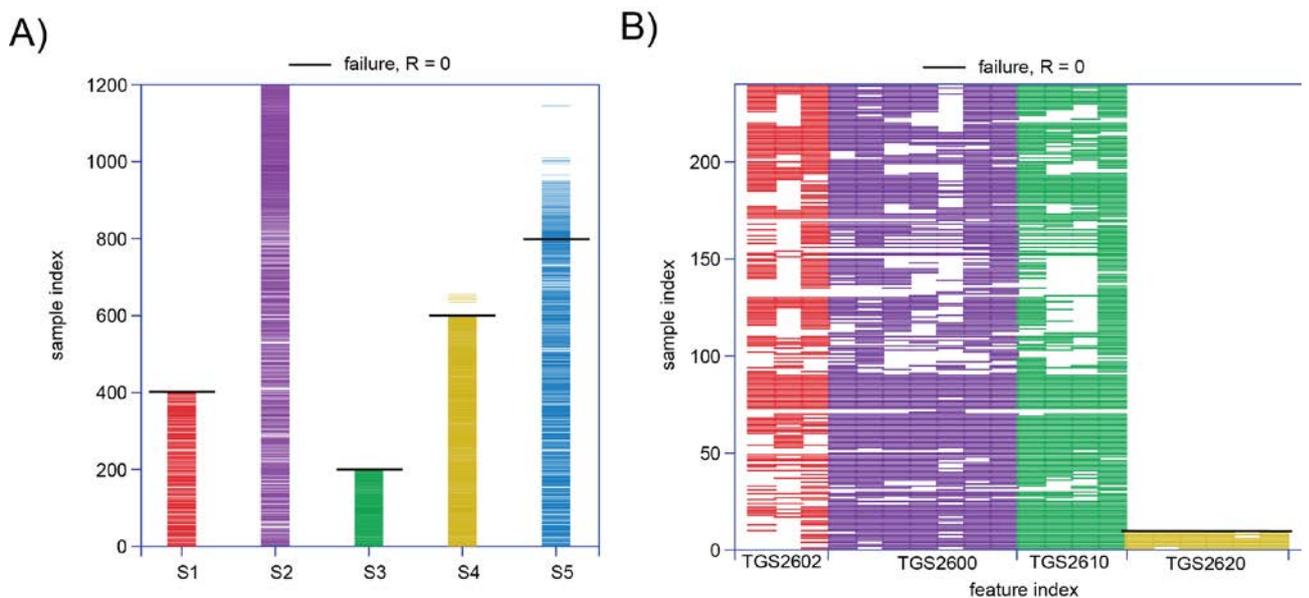

**Figure 9.** Feature selection process during a zero-response failure in case of synthetic and experimental dataset, panel A) and B) respectively. LDA has been used as core classifier.

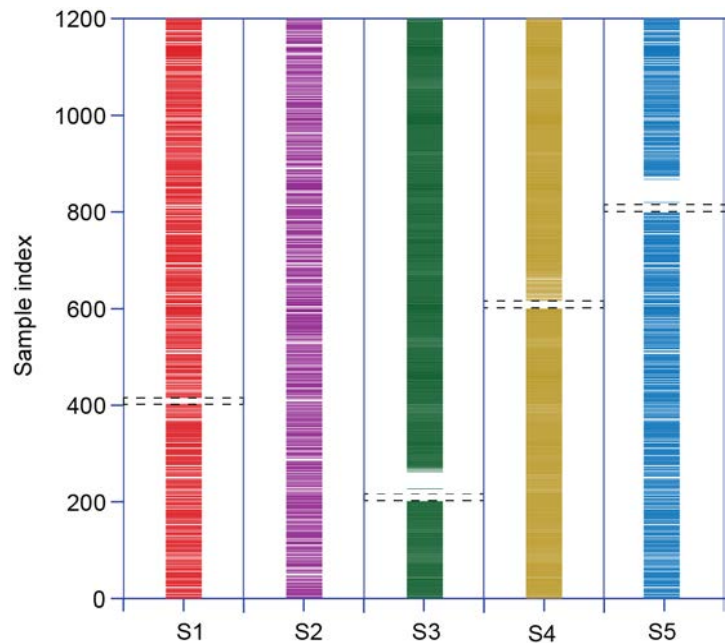

**Figure 10.** Feature selection process during a temporary zero-response faults in case of synthetic dataset. Faults last for 15 samples. LDA is used as core classifier.

## 6. Conclusions

In this paper, we investigated the properties and performances of Self-Repairing, an algorithm that can be used to replace faulty sensor, to improve classification and to restore the original redundancy of an array preserving the fault tolerance. Self-Repairing algorithm operates during the test phase without interrupting the classification procedure. Furthermore, it can implement, thanks to the UOS algorithm, any standard model as core classifier. Results obtained in case of synthetic and experimental datasets showed that the algorithm is robust in case of both long lasting experiment with multiple faults and in case of extremely irregular dataset where classes are presented in blocks rather than alternated. This last case is crucial for adaptive model as investigate in ref [39]. The initial sensor arrays were selected without any replica to show the potentialities of algorithm in a more challenging context. Obviously implementing sensor replicas from the beginning it is possible to obtain even better performances. Finally, the feature selection process can be used itself as a parameter to detect and monitoring the occurrences of sensor failures and faults.

**Acknowledgements**

This work is partially funded by Phasmafood project "PhasmaFOOD: Portable photonic miniaturised smart system for on-the-spot food quality sensing" H2020-ICT-2016-2017 (H2020-ICT-2016-1)

## 7. References.

[1] Banica, F. (2012). Chemical Sensors and Biosensors: Fundamentals and Applications. John Wiley & Sons (UK).

[2] Korotcenkov, G., and B. K. Cho. "Instability of metal oxide-based conductometric gas sensors and approaches to stability improvement (short survey)." Sensors and Actuators B: Chemical 156.2 (2011): 527-538.

[3] Korotcenkov, Ghenadii. "Metal oxides for solid-state gas sensors: What determines our choice?." Materials Science and Engineering: B 139.1 (2007): 1-23.

[4] Romain, A., & Nicolas, J. (2010). Long term stability of metal oxide-based gas sensors for e-nose environmental applications: An overview. Sensors and Actuators B: Chemical, 146(2), 502-506. doi:10.1016/j.snb.2009.12.027

[5] Di Natale, C., Marco, S., Davide, F., & D'amico, A. (1995). Sensor-array calibration time reduction by dynamic modelling. Sensors and Actuators B: Chemical, 25(1-3), 578-583. doi:10.1016/0925-4005(95)85126-7

[6] Homer, M., Shevade, A., Lara, L., Huerta, R., Vergara, A., & Muezzinoglu, M. (2012). Rapid Analysis, Self-Calibrating Array for Air Monitoring. 42nd International Conference on Environmental Systems. doi:10.2514/6.2012-3457

[7] Rodriguez-Lujan, I., Fonollosa, J., Vergara, A., Homer, M., & Huerta, R. (2014). On the calibration of sensor arrays for pattern recognition using the minimal number of experiments. Chemometrics and Intelligent Laboratory Systems, 130, 123-134. doi:10.1016/j.chemolab.2013.10.012

[8] Castro, R. M., & Nowak, R. D. (n.d.). Minimax Bounds for Active Learning. Learning Theory Lecture Notes in Computer Science, 5-19. doi:10.1007/978-3-540-72927-3_3

[9] Poggio, T., Rifkin, R., Mukherjee, S., & Niyogi, P. (2004). General conditions for predictivity in learning theory. Nature, 428(6981), 419-422. doi:10.1038/nature02341

[10] Artursson, T., Eklöv, T., Lundström, I., Mårtensson, P., Sjöström, M., & Holmberg, M. (2000). Drift correction for gas sensors using multivariate methods. Journal of chemometrics, 14(5‐6), 711-723.

[11] Romain, A. C., & Nicolas, J. (2010). Long term stability of metal oxide-based gas sensors for e-nose environmental applications: An overview. Sensors and Actuators B: Chemical, 146(2), 502-506.

[12] Roth, M., Hartinger, R., Faul, R., & Endres, H. (1996). Drift reduction of organic coated gas-sensors by temperature modulation. Sensors and Actuators B: Chemical, 36(1-3), 358-362. doi:10.1016/s0925-4005(97)80096-2

[13] Martinelli, E., Polese, D., Catini, A., D'Amico, A., & Natale, C. D. (2012). Self-adapted temperature modulation in metal-oxide semiconductor gas sensors. Sensors and Actuators B: Chemical, 161(1), 534-541. doi:10.1016/j.snb.2011.10.072

[14] Vergara, A., Martinelli, E., Llobet, E., D'amico, A., & Natale, C. D. (2009). Optimized Feature Extraction for Temperature-Modulated Gas Sensors. Journal of Sensors, 2009, 1-10. doi:10.1155/2009/716316

[15] Hui, D., Jun-Hua, L., & Zhong-Ru, S. (2003). Drift reduction of gas sensor by wavelet and principal component analysis. Sensors and Actuators B: Chemical, 96(1-2), 354-363. doi:10.1016/s0925-4005(03)00569-0

[16] Martinelli, E., Pennazza, G., Natale, C. D., & D'Amico, A. (2004). Chemical sensors clustering with the dynamic moments approach. Sensors and Actuators B: Chemical, 101(3), 346-352. doi:10.1016/j.snb.2004.04.010

[17] Gutierrez-Osuna, Ricardo, and Andreas Hierlemann. "Adaptive microsensor systems." Annual Review of Analytical Chemistry 3 (2010): 255-276.

[18] Di Carlo, S., Falasconi, M., Sánchez, E., Scionti, A., Squillero, G., & Tonda, A. (2010). Exploiting Evolution for an Adaptive Drift-Robust Classifier in Chemical Sensing. Applications of Evolutionary Computation Lecture Notes in Computer Science, 412-421. doi:10.1007/978-3-642-12239-2_43

[19] Vergara, A., Vembu, S., Ayhan, T., Ryan, M. A., Homer, M. L., & Huerta, R. (2012). Chemical gas sensor drift compensation using classifier ensembles. Sensors and Actuators B: Chemical, 166-167, 320-329. doi:10.1016/j.snb.2012.01.074

[20] Padilla, M., Perera, A., Montoliu, I., Chaudry, A., Persaud, K., & Marco, S. (2010). Drift compensation of gas sensor array data by Orthogonal Signal Correction. Chemometrics and Intelligent Laboratory Systems, 100(1), 28-35. doi:10.1016/j.chemolab.2009.10.002

[21] Haugen, J., Tomic, O., & Kvaal, K. (2000). A calibration method for handling the temporal drift of solid state gas-sensors. Analytica Chimica Acta, 407(1-2), 23-39. doi:10.1016/s0003-2670(99)00784-9

[22] Zuppa, M. (2004). Drift counteraction with multiple self-organising maps for an electronic nose. Sensors and Actuators B: Chemical, 98(2-3), 305-317. doi:10.1016/j.snb.2003.10.029

[23] Hines, E., Llobet, E., & Gardner, J. (1999). Electronic noses: A review of signal processing techniques. IEE Proceedings - Circuits, Devices and Systems, 146(6), 297. doi:10.1049/ip-cds:19990670

[24] Reimann, P., Dausend, A., & Schutze, A. (2008). A self-monitoring and self-diagnosis strategy for semiconductor gas sensor systems. 2008 IEEE Sensors. doi:10.1109/icsens.2008.4716415

[25] Pardo, M., Faglia, G., Sberveglieri, G., Corte, M., Masulli, F., & Riani, M. (2000). Monitoring reliability of sensors in an array by neural networks. Sensors and Actuators B: Chemical, 67(1-2), 128-133. doi:10.1016/s0925-4005(00)00402-0

[26] Isermann, R., Balle, P.: Trends in the application of model-based fault detection and diagnosis of technical processes. Control Eng. Pract. 5, 709–719 (1997)

[27] Ni, K., Ramanathan, N., Chehade, M. N. H., Balzano, L., Nair, S., Zahedi, S., ... & Srivastava, M. (2009). Sensor network data fault types. ACM Transactions on Sensor Networks (TOSN), 5(3), 25.

[28] Qin, S. J., & Li, W. (1999). Detection, identification, and reconstruction of faulty sensors with maximized sensitivity. *AIChE journal*, *45*(9), 1963-1976.

[29] Magna, G., Vergara, A., Martinelli, E., & Di Natale, C. (2014). Automatic Fault Identification and On-line Unsupervised Calibration of Replaced Sensors by Means of Cooperative Classifiers. Procedia Engineering, 87, 855-858, doi:10.1016/j.proeng.2014.11.288

[30] Tomic, O., Eklöv, T., Kvaal, K., & Haugen, J. E. (2004). Recalibration of a gas-sensor array system related to sensor replacement. Analytica Chimica Acta, 512(2), 199-206.

[31] Zhang, L., Tian, F., Peng, X., Dang, L., Li, G., Liu, S., & Kadri, C. (2013). Standardization of metal oxide sensor array using artificial neural networks through experimental design. Sensors and Actuators B: Chemical, 177, 947-955.

[32] Solórzano, A., Rodríguez-Pérez, R., Padilla, M., Graunke, T., Fernandez, L., Marco, S., & Fonollosa, J. (2018). Multi-unit calibration rejects inherent device variability of chemical sensor arrays. *Sensors and Actuators B: Chemical*, *265*, 142-154.

[33] Magna, G., Mosciano, F., Martinelli, E., & Di Natale, C. (2018). Unsupervised On-Line Selection of Training Features for a robust classification with drifting and faulty gas sensors. Sensors and Actuators B: Chemical, 258, 1242-1251.

[34] Albert, K. J., Lewis, N. S., Schauer, C. L., Sotzing, G. A., Stitzel, S. E., Vaid, T. P., & Walt, D. R. (2000). Cross-reactive chemical sensor arrays. Chemical reviews, 100(7), 2595-2626.

[35] Koushanfar, Farinaz, Miodrag Potkonjak, and Alberto Sangiovanni-Vincentelli. "On-line fault detection of sensor measurements." *Sensors, 2003. Proceedings of IEEE*. Vol. 2. IEEE, 2003.

[36] Isermann, Rolf. "Model-based fault-detection and diagnosis–status and applications." *Annual Reviews in control* 29.1 (2005): 71-85.

[37] Luo, Rongfu, Manish Misra, and David M. Himmelblau. "Sensor fault detection via multiscale analysis and dynamic PCA." *Industrial & Engineering Chemistry Research* 38.4 (1999): 1489-1495.

[38] Chiang, Leo H., Evan L. Russell, and Richard D. Braatz. "Fault diagnosis in chemical processes using Fisher discriminant analysis, discriminant partial least squares, and principal component analysis." *Chemometrics and intelligent laboratory systems* 50.2 (2000): 243-252.

[39] Martinelli, E., Magna, G., De Vito, S., Di Fuccio, R., Di Francia, G., Vergara, A., & Di Natale, C. (2013). An adaptive classification model based on the Artificial Immune System for chemical sensor drift mitigation. *Sensors and Actuators B: Chemical*, *177*, 1017-1026.

[40] Vergara, A., Vembu, S., Ayhan, T., Ryan, M. A., Homer, M. L., & Huerta, R. (2012). Chemical gas sensor drift compensation using classifier ensembles. *Sensors and Actuators B: Chemical*, *166*, 320-329.